\documentclass[aps,prd,a4paper,preprintnumbers,superscriptaddress,twocolumn,floatfix,showpacs,nofootinbib]{revtex4}
\usepackage{amssymb}
\usepackage{amsmath}

\usepackage{float,graphicx}

\begin{document}

\title{Does Bulk Viscosity Create a Viable Unified Dark Matter Model?}
\author{Baojiu~Li}
\email[Email address: ]{b.li@damtp.cam.ac.uk}
\affiliation{DAMTP, Centre for Mathematical Sciences, University of Cambridge, Cambridge
CB3 0WA, United Kingdom}
\author{John~D.~Barrow}
\email[Email address: ]{j.d.barrow@damtp.cam.ac.uk}
\affiliation{DAMTP, Centre for Mathematical Sciences, University of Cambridge, Cambridge
CB3 0WA, United Kingdom}
\date{\today}

\begin{abstract}
We investigate in detail the possibility that a single imperfect fluid with
bulk viscosity can replace the need for separate dark matter and dark energy
in cosmological models. With suitable choices of model parameters, we show
that the background cosmology in this model can mimic that of a $\Lambda $%
CDM Universe to high precision. However, as the cosmic expansion
goes through the decelerating-accelerating transition, the density
perturbations in this fluid are rapidly damped out. We show that,
although this does not significantly affect structure formation in
baryonic matter, it makes the gravitational potential decay
rapidly at late times, leading to modifications in predictions of
cosmological observables such as the CMB power spectrum and weak
lensing. This model of unified dark matter is thus difficult to
reconcile with astronomical observations. We also clarify the
differences with respect to other unified dark matter models where
the fluid is barotropic, \emph{i.e.}, $p=p(\rho )$, such as the
(generalized) Chaplygin gas model, and point out their
observational difficulties. We also summarize the status of dark
sector models with no new dynamical degrees of freedom introduced
and discuss the problems with them.
\end{abstract}


\maketitle

\section{Introduction}

\label{sect:Introduction}

In recent years the cosmological picture that over 95\% of the energy in our
universe is contributed by a dark sector has been supported independently by
a number of observations, notably those of type Ia Supernovae (SN)
luminosity distances, cosmic microwave background (CMB) anisotropies, the
power spectrum of clustered matter, and weak lensing \cite{SN, Riess2004,
CMB, Pk}. This dark sector can be further subdivided into dark matter and
dark energy according to their different gravitational properties. The
concordance $\Lambda $CDM paradigm, in which dark matter is assumed to be
weakly (or just gravitationally) interacting massive particles (the cold
dark matter), and dark energy is a positive cosmological constant ($\Lambda $%
) or slowly varying scalar field, has been successful in confronting all
these observational data sets. However, the smallness of the cosmological
constant, and the fact that it only becomes dominant recently make this
model conceptually unattractive and stimulates the examination of new models
(for a review see \cite{DEReview}).

Since both cold dark matter (CDM) and dark energy are invisible
and have as yet unknown origins, it is natural to consider the
possibility that they are actually not two exotic matter species
but just different aspects of a single fluid. These scenarios are
frequently dubbed unified dark matter (UDM) models. In the context
of general relativity they assume an equation of state $p=p(\rho
)$ or $p=p(H)$, where $H$ is the Hubble rate; in universes with
zero spatial curvature these prescriptions are identical. A
particular class of cosmology with this equation of state was
investigated in the context of studies of cosmological bulk
viscosity \cite{Bel, Barrow1988, Szydlowski2007, Pavon1991,
Pavon1993, Pavon2001} in which the viscosity coefficient is $\eta
(\rho )=\alpha \rho ^{m}$ and the effective pressure in the
Friedmann equation is $p^{\prime
}=(\gamma -1)\rho -3H\eta (\rho )$, and of string production effects \cite%
{turok} which mimic the effects of bulk viscosity of this form with $m=3/2$
\cite{Barrow1988}.

The flat bulk viscous cosmologies also include as subcases the so called
Chaplygin gas model ($p=-A\rho ^{-\alpha }$) \cite{JDB1990, Kamenshchik2001,
Bento2002}, which is just a bulk viscosity for dust ($\gamma =1$) with $%
m+1/2\equiv -\alpha $, and its generalizations to $\rho +p=B\rho ^{\lambda }$
\cite{JDB1990}, which is just $\gamma =0$ and $m+1/2\equiv \lambda $, or
other functional forms $p(\rho )$ \cite{DEReview}. The Chaplygin gas models
are simple, with no new dynamical degrees of freedom, and yet produce an
interesting time-evolution for the dark energy. However, it was shown
subsequently that the density perturbation of this fluid will either blow up
or experience rapid oscillatory damping at late times, so the models can be
stringently constrained by the matter power spectrum \cite{Sandvik2004}.
This distinctive behaviour arises because there is a minimum possible total
density of the universe at late times.

There have been some explicit investigations in the bulk viscous cosmology
in connection with the dark energy problem. Fabris \emph{et~al}.~\cite%
{Fabris2006, Colistete2007} investigated the case in which the viscosity
coefficient has the form $\eta (\rho )=\alpha \rho ^{m}$ and considered both
background and perturbed evolutions of a universe dominated by viscous
matter. However, these studies were limited by certain simplifications. In
Ref.~\cite{Fabris2006}, for example, the authors obtained analytical
solution for $\rho _{v}$ (where the subscript $_{v}$ denotes viscous matter)
as a function of scale factor $a$, under the assumption that no other matter
species exist in the universe (the same assumption as \cite{Barrow1988}). In
Ref.~\cite{Colistete2007} the authors added a baryonic matter species and
estimated the cosmological parameters in detail using Bayesian statistics;
but their calculation was largely confined to the special case $m=0$ and
used only supernovae data to derive the constraints. We believe that such an
analysis needs to be generalized. First, as will be seen below, $m=0$ is not
necessarily the best fit for these models, even for the background
cosmology. Second, no radiation component was included in previous analysis.
This is a reasonable simplification as long as we are only concerned with
the late-time cosmic expansion and only using the supernova data. But when
observables are also related to the high-redshift features, such as the CMB
shift parameter, the radiation must be taken into account. In fact, the
properties of the viscosity could significantly modify the early-time
evolution of the universe if the late-time evolution is consistent with the
supernova data, and so the CMB shift parameter could be effective in
constraining the model. Third, a detailed study of the linear perturbations
in the model is needed as a critical check of the model's feasibility. As we
will see below, the UDM model based on viscous matter has several
distinctive predictions regarding the perturbations when compared to the $%
\Lambda $CDM paradigm or the Chaplygin gas model. Such a feature is very
general, making the model barely compatible with observations. This
indicates that a UDM model without new dynamical degrees of freedom is
unlikely to be observationally acceptable.

The rest of this paper is organized as follows. In \S ~\ref{sect:Equations},
we give the basic field equations which will be used in the subsequent
analysis. In \S ~\ref{sect:BackgroundEvolution} we consider the background
evolution of a universe dominated by a viscous matter (in the presence of
normal matter species such as baryons, photons and neutrinos). Using the SN
data and CMB shift parameter, we find the best-fit model parameters and show
that these give a background cosmology very similar to the prediction of the
$\Lambda $CDM paradigm. The \S ~\ref{sect:PerturbedEvolution} is devoted to
the perturbed evolution of general bulk viscosity models, which we find to
behave very differently from both $\Lambda $CDM and other UDM models.
Finally, we discuss our results and conclude in \S ~\ref{sect:conclusion}.
We will frequently call the UDM fluid with $p=p(\rho ,H)$ 'the viscous dark
matter'.

\section{The Field Equations}

\label{sect:Equations}

In this section we list the general field equations that govern
the evolution of the cosmological background and its first-order
perturbations in general relativity, which will be used in later
sections. The perturbation equations will be given in the
covariant and gauge invariant (CGI) formalism, using the method of
$3+1$ decomposition.

The main idea of $3+1$ decomposition is to make spacetime splits
of physical quantities with respect to the 4-velocity $u^{a}$ of
an observer. The projection tensor $h_{ab}$ is defined as
$h_{ab}=g_{ab}-u_{a}u_{b}$ and can be used to obtain covariant
tensors perpendicular to $u$. For example, the covariant spatial
derivative $\hat{\nabla}$ of a tensor field $T_{d\cdot \cdot \cdot
e}^{b\cdot \cdot \cdot c}$ is defined as
\begin{eqnarray}  \label{eq:AEEOM}
\hat{\nabla}^{a}T_{d\cdot \cdot \cdot e}^{b\cdot \cdot \cdot
c}\equiv h_{i}^{a}h_{j}^{b}\cdot \cdot \cdot \
h_{k}^{c}h_{d}^{r}\cdot \cdot \cdot \ h_{e}^{s}\nabla
^{i}T_{r\cdot \cdot \cdot s}^{j\cdot \cdot \cdot k}.
\end{eqnarray}%
The energy-momentum tensor and covariant derivative of the
4-velocity are decomposed respectively as
\begin{eqnarray}  \label{eq:AEEMT}
T_{ab} &=& \pi _{ab}+2q_{(a}u_{b)}+\rho u_{a}u_{b}-ph_{ab}, \\
\nabla _{a}u_{b} &=&\sigma _{ab}+\varpi _{ab}+\frac{1}{3}\theta
h_{ab}+u_{a}A_{b}.
\end{eqnarray}
In the above, $\pi_{ab}$ is the projected symmetric trace-free
(PSTF) anisotropic stress, $q_{a}$ the vector heat flux vector,
$p$ the isotropic pressure, $\sigma_{ab}$ the PSTF shear tensor,
$\varpi_{ab}=\hat{\nabla}_{[a}u_{b]}$, the vorticity, $\theta
=\nabla^{c}u_{c}\equiv 3\dot{a}/a$ ($a$ is the mean expansion
scale factor) the expansion scalar, and $A_{b}=\dot{u}_{b}$ the
acceleration; the overdot denotes time derivative expressed as
$\dot{\phi}=u^{a}\nabla _{a}\phi $, brackets mean
antisymmetrisation, and parentheses symmetrization. The 4-velocity
normalization is chosen to be $u^{a}u_{a}=1$. The quantities
$\pi_{ab}, q_{a}, \rho, p$ are referred to as \emph{dynamical}
quantities and $\sigma _{ab}, \varpi _{ab}, \theta, A_{a}$ as
\emph{kinematical} quantities. Note that the dynamical quantities
can be obtained from the energy-momentum tensor $T_{ab}$ through
the relations
\begin{eqnarray}  \label{eq:DefDynamicalQuantity}
\rho &=&T_{ab}u^{a}u^{b},  \notag \\
p &=&-\frac{1}{3}h^{ab}T_{ab},  \notag \\
q_{a} &=&h_{a}^{d}u^{c}T_{cd},  \notag \\
\pi _{ab} &=&h_{a}^{c}h_{b}^{d}T_{cd}+ph_{ab}.
\end{eqnarray}

Decomposing the Riemann tensor and making use the Einstein
equations, we obtain, after linearization, a set of propagation
and constraint equations governing the evolution of perturbed
physical quantities. Here we shall only list the equations that
will be used in later sections, and for more details we refer the
reader to \cite{GR3+1}.

The first equation we will use is the Raychaudhuri equation
\begin{eqnarray}\label{eq:raychaudhrui}
\dot{\theta}+\frac{1}{3}\theta^{2}-\hat{\nabla}^{a}A_{a}+\frac{\kappa
}{2}(\rho+3p) &=& 0.
\end{eqnarray}

The second equation to be used involves the projected Ricci scalar
$\hat{R}$ into the hypersurfaces orthogonal to $u^{a}$, which can
be expressed as
\begin{eqnarray}\label{eq:SpatialRicciCurvature}
\hat{R} &\doteq& 2\kappa \rho -\frac{2}{3}\theta ^{2}.
\end{eqnarray}
Since we are considering a spatially-flat universe, the spatial
curvature vanishes (at background level) on large scales and so
$\hat{R}=0$. Thus, from Eq.~(\ref{eq:SpatialRicciCurvature}), we
obtain
\begin{eqnarray}
\frac{1}{3}\theta ^{2} &=& \kappa \rho ,
\end{eqnarray}
which is the Friedmann equation that governs the expansion of the
universe in standard general relativity.

Furthermore, we will need the conservation equations for the
energy-momentum tensor
\begin{eqnarray}
\label{eq:EnergyConservation}\dot{\rho}+(\rho +p)\theta
+\hat{\nabla}^{a}q_{a} &=& 0,\\
\label{eq:conserv_q} \dot{q}_{a}+\frac{4}{3}\theta q_{a}+(\rho
+p)A_{a}-\hat{\nabla}_{a}p+\hat{\nabla}^{b}\pi_{ab} &=& 0.
\end{eqnarray}

Eqs.~(\ref{eq:raychaudhrui}, \ref{eq:EnergyConservation}, \ref{eq:conserv_q}%
) involves both background and first-order perturbed quantities (such as $%
A_{a}$, $q_{a}$). To obtain equations for the background cosmology
it is sufficient to neglect all first-order terms, while to obtain
corresponding equations for the perturbation evolution we need to
take the spatial derivatives $\hat{\nabla}_{a}$
[cf.~Eq.~(\ref{eq:AEEOM})] of these equations. For the
perturbation analysis it is then more convenient to work in the
$k$ space because we shall confine ourselves to the linear regime
where different $k$-modes decouple. Following \cite{GR3+1}, we
make the following harmonic expansions of our perturbation
variables:
\begin{eqnarray} \label{eq:HarmonicExpansion}
\hat{\nabla}_{a}\rho = \sum_{k}\frac{k}{a}\mathcal{X}Q_{a}^{k}\ \
\ \ \hat{\nabla}_{a}p =
\sum_{k}\frac{k}{a}\mathcal{X}^{p}Q_{a}^{k}\nonumber\\ q_{a} =
\sum_{k}qQ_{a}^{k}\ \ \ \ \pi_{ab} = \sum_{k}\Pi Q_{ab}^{k}\ \ \ \
\nonumber\\
\hat{\nabla}_{a}\theta =
\sum_{k}\frac{k^{2}}{a^{2}}\mathcal{Z}Q_{a}^{k}\ \ \ \ A_{a} =
\sum_{k}\frac{k}{a}AQ^{k}_{a}\nonumber
\end{eqnarray}
in which $Q^{k}$ is the eigenfunction of the comoving spatial
Laplacian $a^{2}\hat{\nabla}^{2}$ satisfying
\begin{eqnarray}
\hat{\nabla}^{2}Q^{k} &=& \frac{k^{2}}{a^{2}}Q^{k},
\end{eqnarray}
and $Q_{a}^{k},Q_{ab}^{k}$ are given by
$Q_{a}^{k}=\frac{a}{k}\hat{\nabla}_{a}Q^{k},Q_{ab}^{k}=\frac{a}{k}\hat{\nabla}_{\langle
a}Q_{b\rangle }^{k}$.

The perturbed version of Eq.~(\ref{eq:EnergyConservation}) is
\begin{eqnarray}  \label{eq:conserv_rho}
\dot{\mathcal{X}}_{a} + (\rho+p)(\mathcal{Z}_{a} - a\theta A_{a})
+ (\mathcal{X}_{a} + \mathcal{X}^{p}_{a})\theta +
a\hat{\nabla}_{a}\hat{\nabla}^{b}q_{b} &=& 0.
\end{eqnarray}
In \S ~\ref{sect:PerturbedEvolution} we shall use the harmonic
expansion coefficients of Eqs.~(\ref{eq:raychaudhrui},
\ref{eq:conserv_q}, \ref{eq:conserv_rho}) to derive the evolution
equation for the density perturbation of the viscous dark matter.

\section{The Background Evolution}

\label{sect:BackgroundEvolution}

The field equations governing the background cosmic expansion are the
Friedmann equation, the Raychaudhuri equation and the conservation equations
for energy densities of the different matter species (baryons, radiation,
and viscous dark matter). Not all these equations are independent, and we
choose the Friedmann equation (here $H=\dot{a}/a=\theta /3$)
\begin{eqnarray} \label{eq:Friedman}
3H^{2} &=& 8\pi G\left(\rho_{\mathrm{D}} + \rho_{\mathrm{B}} +
\rho_{\mathrm{R}}\right),
\end{eqnarray}
and the conservation equations
\begin{eqnarray}
\label{eq:conserv_b} \dot{\rho}_{\mathrm{B}} + 3H\rho_{\mathrm{B}} &=& 0,\\
\label{eq:conserv_r} \dot{\rho}_{\mathrm{R}} + 4H\rho_{\mathrm{R}} &=& 0,\\
\label{eq:conserv_vdm} \dot{\rho}_{\mathrm{D}} +
3H\left(\rho_{\mathrm{D}} + p_{\mathrm{D}}\right) &=& 0
\end{eqnarray}
as our starting point. In these equations $\rho _{\mathrm{B}},
\rho _{\mathrm{R}}$ and $\rho_{\mathrm{D}}$ are respectively the
energy densities of baryonic matter, radiation and the (viscous)
dark matter. The pressure
$p_{\mathrm{D}}=(\gamma-1)\rho_{\mathrm{D}}-3\alpha
H\rho_{\mathrm{D}}^{m}$ is the effective pressure of the dark
matter, with $\gamma, \alpha, m$ being our free model parameters.

We shall not follow \cite{Colistete2007} by dividing
$\rho_{\mathrm{D}}$ into different components with different
equation of states (EoS), although it can be useful
mathematically, since that might hide the fact that there is only
a single fluid with varying EoS. When it comes to the perturbative
evolution of this fluid, it will be misleading if one thinks that
part of the fluid behaves as a cosmological constant which has no
perturbation, while another part simply clusters as CDM, there
being \emph{no} interactions between them. We also note that the
Hubble expansion rate $H$ appears in the expression of
$p_{\mathrm{D}}$, which means that the EoS of the viscous dark
matter depends also on the existence and properties of other
matter species, and so we cannot neglect the effects of baryonic
matter and radiation (at early times).

\begin{figure}[tbp]
\centering \includegraphics[scale=.8] {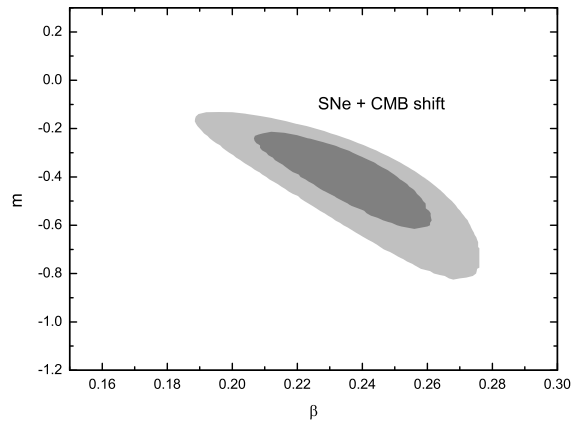}
\caption{The joint constraints on $m$ and $\protect\beta $ from Supernovae
and CMB shift-parameter data. The dark grey and light grey areas denote the
68\% and 95\% confidence regions respectively. The current Hubble expansion
rate $H_{0}$ is marginalized over analytically and the other parameters used
are $\Omega _{\mathrm{R}}=8.475\times 10^{-5}$ and $\Omega _{\mathrm{R}%
}+\Omega _{\mathrm{B}}=0.04$ evaluated at the present time.}
\label{fig:Figure1}
\end{figure}

Next, we estimate the free model parameters $\gamma ,\alpha $ and
$m$. We will set $\gamma =1$ in all the calculations below because
we want the viscous dark matter to behave like CDM at (early)
times when the correction term is not important. For $m$, earlier
studies \cite{Barrow1988} showed that when $m>1/2$ the universe
will start from a de Sitter phase and finally evolve towards
power-law perfect-fluid dominated expansion, while if $m<1/2$ it
is just the opposite, with late-time approach to de Sitter
evolution; the special $m=1/2$ case corresponds to power law
evolution throughout. In fact, from the expression
$p_{\mathrm{D}}=-3\alpha H\rho _{\mathrm{D}}^{m}$ we can see that
if $\rho _{\mathrm{D}}$ dominates over other matter species (so
that $H\propto \rho _{\mathrm{D}}^{1/2}$) then $p_{\mathrm{D}}$
will be a constant if $m=-1/2$. We can estimate that the best fit
value of $m$ is around $-1/2$. In what follows, $m$ will be taken
as a free parameter to be constrained by data. Finally, $\alpha $
is obviously a dimensional constant. To determine its value, we
note that to explain the SNe data we need the candidate for dark
energy to contribute an effective energy density and pressure of
the same order as $\rho _{\Lambda }$ in the $\Lambda $CDM model.
Thus, we have
\begin{eqnarray}
\rho_{\mathrm{D}0} &\simeq& \rho_{\mathrm{CDM}0} + \rho_{\Lambda},
\nonumber\\
3\alpha H_{0}\rho^{m}_{\mathrm{D}0} &\simeq&
\rho_{\Lambda}\nonumber
\end{eqnarray}
in which the quantities on the left-hand sides are for our model
while those on the right hand sides are for the $\Lambda$CDM
model. A subscript $_{0}$ here denotes the present value of a
quantity. Using the results $\Omega_{\mathrm{CDM}}\simeq 0.20$ and
$\Omega_{\Lambda}\simeq 0.76$, we calculate from the above two
equations that $\beta \equiv \alpha
H_{0}\rho_{\mathrm{DM}0}^{m-1}\simeq 0.26$. This gives us a sense
about the magnitude of the dimensionless quantity $\beta$, which
is chosen as another model parameter instead of $\alpha$ and will
be constrained below.

With the above preliminaries, it is now straightforward to rewrite
the conservation equation of the viscous dark matter,
Eq.~(\ref{eq:conserv_vdm}), as
\begin{eqnarray} \label{eq:conserv_vdm_2}
\varrho^{\ast} + 3\gamma\varrho &=&
9\beta\varrho^{m}\left[\frac{\varrho + r_{bd}e^{-3N} +
r_{rd}e^{-4N}}{1+r_{bd}+r_{rd}}\right]^{1/2}
\end{eqnarray}
where we have defined $\varrho \equiv
\rho_{\mathrm{DM}}/\rho_{\mathrm{DM} 0}$, $r_{bd}\equiv
\rho_{\mathrm{B}0}/\rho_{\mathrm{DM}0}$, $r_{rd}\equiv
\rho_{\mathrm{R}0}/\rho_{\mathrm{DM}0}$, and also used the
Friedmann equation Eq.~(\ref{eq:Friedman}) to substitute for the
Hubble parameter, $H$. The star denotes a derivative with respect
to $N=\log a$. Because there are in general no closed-form
solutions to these cosmological equations, we will use
Eq.~(\ref{eq:conserv_vdm_2}), together with
Eqs.~(\ref{eq:Friedman}, \ref{eq:conserv_b}, \ref{eq:conserv_r}),
in our subsequent numerical calculation. Note that $r_{bd}$ and
$r_{rd}$ are constants which are fixed once $\Omega_{\mathrm{B}}$
and $\Omega _{\mathrm{R}}$ are known (using the fact that the
universe is spatially flat, so that
$\Omega_{\mathrm{DM}}=1-\Omega_{\mathrm{B}}-\Omega_{\mathrm{R}}$):
$\Omega_{\mathrm{DM}}=1-\Omega_{\mathrm{B}}-\Omega_{\mathrm{R}}$):
\begin{eqnarray}
r_{bd}\ =\ \frac{\Omega_{\mathrm{B}}}{1 - \Omega_{\mathrm{B}} -
\Omega_{\mathrm{R}}},\ \ \ r_{rd}\ =\ \frac{\Omega_{\mathrm{R}}}{1
- \Omega_{\mathrm{B}} - \Omega_{\mathrm{R}}}.
\end{eqnarray}
A natural choice of the initial (final) condition of
Eq.~(\ref{eq:conserv_vdm_2}) is $\varrho (N=0)=1$.

\begin{figure}[tbp]
\centering \includegraphics[scale=.85] {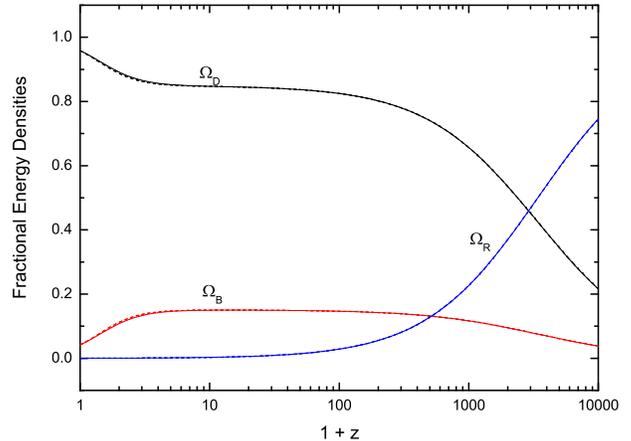}
\caption{(Color Online) Solid lines: The evolution of the fractional energy
densities of viscous dark matter $\Omega _{\mathrm{D}}$ (black), baryons $%
\Omega _{\mathrm{B}}$ (red), and radiation (including photons and massless
neutrinos) $\Omega _{\mathrm{R}}$ (blue), versus $1+z$ where $z$ is the
redshift. The model parameters here are chosen as $m=-0.4$ and $\protect%
\beta =0.236$. Dashed lines: the same evolutions for the concordance $%
\Lambda $CDM model. Here, $\Omega _{\mathrm{D}}$ denotes the fractional
energy density of the dark sector, \emph{i.e.}, dark energy (a cosmological
constant) plus cold dark matter. The other parameters used (for both models)
are $\Omega _{\mathrm{R}}=8.475\times 10^{-5}$ and $\Omega _{\mathrm{R}%
}+\Omega _{\mathrm{B}}=0.04,$ evaluated at the present time.}
\label{fig:Figure2}
\end{figure}

We have used the supernovae luminosity distance data
\cite{Riess2004} and CMB shift parameter \cite{Wang2006} to
constrain the model parameters $m$ and $\beta$, analytically
marginalizing the current Hubble expansion rate $H_{0}$ and
assuming the baryon density today as fixed by BBN and measurements
of the light element abundances. The result is shown in
Fig.~\ref{fig:Figure1} and the best-fit parameters we find are
$(m,\beta )=(-0.4,0.236)$ which lie close to our estimate above
($m=-0.5$, $\beta =0.264$). Fig.~\ref{fig:Figure2} shows the
cosmic evolution of the fractional energy densities in the bulk
viscous model with the above best-fit parameters It can be seen
there that the viscous model mimics the concordance $\Lambda $CDM
paradigm extremely well all through its cosmic history. The model
therefore appears to work well as a description of the background
cosmology.

\section{The Evolution of First-Order Density Perturbations}

\label{sect:PerturbedEvolution}

Despite the excellent coincidence between the viscous dark matter and $%
\Lambda $CDM models for the background evolution found in the last section,
we should also investigate the formation of large-scale structure to test
whether the viscous model is also a feasible model for dark energy. In this
section we show that, generally, a bulk viscosity depending on the energy
density,\emph{\ i.e.}, $p_{\mathrm{D}}=-\eta (\rho _{\mathrm{D}})\nabla
^{a}u_{a}$, will significantly influence the formation and evolution of
large-scale cosmological structure. This very different prediction from that
of the $\Lambda $CDM model indicates that stringent constraints can be
placed on the viability of the bulk viscosity models using observational
data on the CMB spectrum, matter power spectrum, and weak gravitational
lensing.

\begin{figure}[tbp]
\centering \includegraphics[scale=.95] {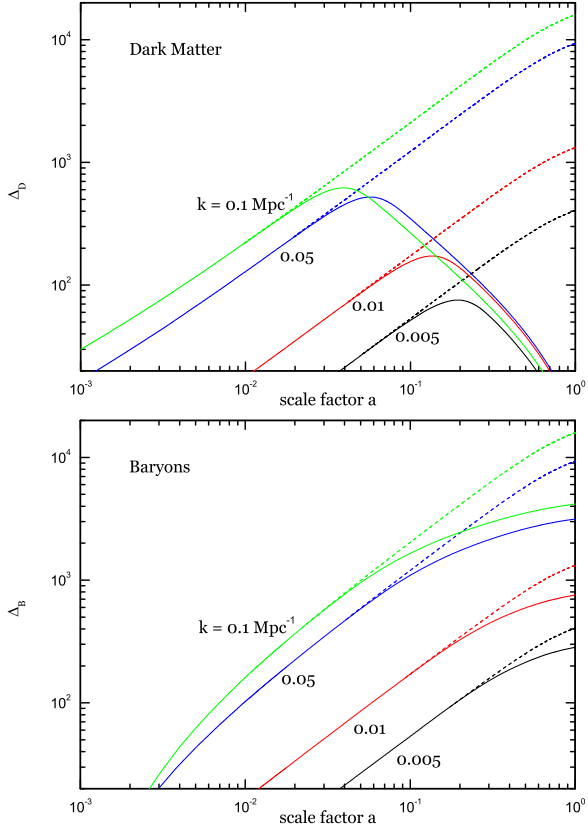}
\caption{(Color Online) \emph{Upper panel}: The evolution of the
density perturbation $\Delta _{\mathrm{D}}$ for the viscous dark
matter (the solid curves), as compared to the predictions for the
cold dark matter model (the
dashed curves). The results for 4 different length scales ($%
k=0.005,0.01,0.05,0.1~\mathrm{Mpc}^{-1}$) are displayed as
labelled beside
the curves. The model parameters are chosen to be ($m=-0.4$, $\protect\beta %
=0.236$). \emph{Lower panel}: The same but for the baryon density
perturbation.} \label{fig:Figure3}
\end{figure}

Let us first concentrate on the special case considered above, with $p_{%
\mathrm{D}}=-\alpha \rho _{\mathrm{D}}^{m}\nabla ^{a}u_{a}=-3\alpha H\rho _{%
\mathrm{D}}^{m}$. For an observer comoving with dark matter particles (with
4-velocity $u_{a}$), the energy momentum tensor could be written as
\begin{eqnarray}\label{eq:EMT}
T_{ab} &=& \rho_{\mathrm{D}}u_{a}u_{b} - p_{\mathrm{D}}(g_{ab} -
u_{a}u_{b}).
\end{eqnarray}
In the $\Lambda $CDM paradigm, if one chooses the observer to be
comoving with the dark matter particles as above, then obviously
the peculiar velocity is zero, $v_{\mathrm{D}}=0$, which implies,
by the conservation of energy-momentum tensor, that $A_{a}=0$ for
this observer, and there is no
acceleration. In the bulk viscous model, however, $v_{\mathrm{D}}=0$ and $%
A_{a}=0$ are different choices of frame which can be used in numerical
calculations. Here, for convenience, we will choose the $v_{\mathrm{D}}=0$
frame, in which, again from the conservation of energy, $A_{a}\neq 0$, which
is easy to understand because the dark-matter particles themselves have
interactions (the pressure $p_{\mathrm{D}}$) and cannot be acceleration-free.

Taking the spatial derivative of $p_{\mathrm{D}}$ and picking the harmonic
coefficients, we obtain
\begin{eqnarray}\label{eq:a_clxcp}
\mathcal{X}_{\mathrm{D}}^{p} &=&
mp_{\mathrm{D}}\Delta_{\mathrm{D}} -
\alpha\rho^{m}_{\mathrm{D}}\frac{k}{a}\mathcal{Z}\ =\
p_{\mathrm{D}}\left(m\Delta_{\mathrm{D}} +
\frac{k\mathcal{Z}}{3\mathcal{H}}\right)
\end{eqnarray}
where $\mathcal{H}=a^{\prime }/a$ with $^{\prime }\equiv d/d\tau $
($\tau$ is the conformal time defined by $ad\tau =dt$) and
$\Delta_{\mathrm{D}}= \mathcal{X}_{\mathrm{D}}/\rho_{\mathrm{D}}$
is the density contrast of the viscous dark matter.

As the anisotropic stress vanishes up to first order in perturbations, from
Eq.~(\ref{eq:conserv_q}) we obtain
\begin{eqnarray}\label{eq:a_A}
(\rho_{\mathrm{D}}+p_{\mathrm{D}})A &=&
\mathcal{X}_{\mathrm{D}}^{p}.
\end{eqnarray}
Similarly, from Eq.~(\ref{eq:conserv_rho}) we have
\begin{eqnarray}\label{eq:a_Delta'}
\Delta'_{\mathrm{D}} + (1+w)k\mathcal{Z} -
3w\mathcal{H}\Delta_{\mathrm{D}} &=& 0
\end{eqnarray}
where we have defined the zero-order EoS parameter $w\equiv w(a)=p_{\mathrm{D%
}}/\rho _{\mathrm{D}}$ for the viscous dark matter so that
\begin{eqnarray}\label{eq:a_w}
w &=& -3\alpha H\rho^{m-1}_{\mathrm{D}}.
\end{eqnarray}
Finally, taking the spatial derivative of the Raychaudhuri equation Eq.~(\ref%
{eq:raychaudhrui}), we get
\begin{eqnarray}\label{eq:a_pert_ray}
k\mathcal{Z}' + k\mathcal{HZ} - k^{2}A + 3(\mathcal{H}' -
\mathcal{H}^{2})A &=& -\frac{\kappa}{2}(\mathcal{X} +
3\mathcal{X}^{p})a^{2}.\ \ \
\end{eqnarray}

We shall assume here that the universe is dominated by the viscous dark
matter, so that on the right-hand side of Eq.~(\ref{eq:a_pert_ray}) $%
\mathcal{X}$ and $\mathcal{X}^{p}$ can be replaced by $\mathcal{X}_{\mathrm{D%
}}$ and $\mathcal{X}_{\mathrm{D}}^{p},$ respectively. In more general cases
it is straightforward to include contributions from other matter species.
Then, from Eqs.~(\ref{eq:a_clxcp}), (\ref{eq:a_A}), (\ref{eq:a_Delta'}) and (%
\ref{eq:a_pert_ray}), we can eliminate $\mathcal{Z}$ and $A$ to obtain
\begin{eqnarray}\label{eq:a_evolve_clxc}
\Delta''_{\mathrm{D}} + \left[C_{1}(a) +
k^{2}C_{2}(a)\right]\Delta'_{\mathrm{D}} + \left[C_{3}(a) +
k^{2}C_{4}(a)\right]\Delta_{\mathrm{D}} &=& 0,
\end{eqnarray}
where we have defined
\begin{eqnarray}\label{eq:C1}
C_{1}(a) &=& \mathcal{H} +
\frac{\kappa\rho\mathrm{_{D}}a^{2}}{2\mathcal{H}}w +
\left(\frac{\mathcal{H}'}{\mathcal{H}} -
\mathcal{H}\right)\frac{w}{1+w}\nonumber\\
&& - \frac{w'}{1+w} - 3w\mathcal{H};\\
\label{eq:C2}C_{2}(a) &=& -\frac{w}{3(1+w)}\frac{1}{\mathcal{H}};\\
\label{eq:C3}C_{3}(a) &=& -\frac{\kappa\rho_{\mathrm{D}}a^{2}}{2}
\left[(1+w)(1+3mw)+3w^{2}\right]\nonumber\\
&& - 3\left(\frac{w'}{1+w}\mathcal{H}+w\mathcal{H}'\right) -
3w\mathcal{H}^{2}\nonumber\\
&& - 3w(\mathcal{H}'-\mathcal{H}^{2})
\left(m+\frac{w}{1+w}\right);\\
\label{eq:C4}C_{4}(a) &=& w\left(m+\frac{w}{1+w}\right).
\end{eqnarray}
It is straightforward to check that when $w=0$ (whether or not $m=0$), this
reduces to the evolution equation for the CDM density contrast, which does
not depend on $k$, indicating that the evolution of the CDM density contrast
is scale-independent. For the viscous dark matter, however, we see that the
equation depends on $k$ in the coefficients of both $\Delta ^{\prime }$ and $%
\Delta $. Eq.~(\ref{eq:a_evolve_clxc}) is effectively an equation for a
damped simple oscillator, with time-varying and scale-dependent frequency
and damping force, and no driving force. On small scales ($k\gg \mathcal{H}$%
) and at late times (when $w\sim \mathcal{O}(-1)$), the $C_{2}(a)$ and $%
C_{4}(a)$ terms dominate the coefficients of $\Delta ^{\prime }$ and $\Delta
,$ respectively; it is then easy to show that the oscillator is over-damped
so that its amplitude decays to zero rapidly with no oscillations. This
qualitative feature can be seen in the upper panel of Fig.~\ref{fig:Figure3}%
, where we plot the evolution of the dark matter density contrast, $\Delta _{%
\mathrm{D}},$ for four different length scales, $k=0.005,0.01,0.05,0.1~%
\mathrm{Mpc}^{-1}$, for our best-fit model parameters ($m=-0.4,\beta =0.236$%
). We see that the evolution of $\Delta _{\mathrm{D}}$ deviates from that
predicted by the $\Lambda $CDM model only at late times, but nonetheless
significantly. The scale-dependent evolution of density contrast is actually
a general feature in the UDM models without new dynamical degrees of freedom
\cite{Sandvik2004, Bean2003, Carturan2003}

The lower panel of Fig.~\ref{fig:Figure3} displays the evolution
of the baryon density contrast $\Delta _{\mathrm{B}}$. Because
viscous dark matter clusters more weakly than cold dark matter,
the gravitational potential that the baryons lie in is also
weaker, making the baryons less clustered. However, since this
effect is indirect, the deviation of $\Delta_{\mathrm{B}}$ from
the $\Lambda $CDM prediction is considerably smaller than that of
$\Delta_{\mathrm{D}}$. Note that the galaxy surveys actually
measure the clustering of luminous (baryonic) matter, rather than
that of dark matter. Consequently, these measurements can only be
applied to $\Delta _{\mathrm{B}} $, which is just weakly dependent
on the model parameters \cite{Beca2003}. Nevertheless, as in the
bulk viscous model, both the amplitude and the shape of baryonic
matter power spectrum are distinct from those predicted by the
$\Lambda$CDM paradigm, and therefore we expect that stringent
constraints can be obtained from it.

\begin{figure}[tbp]
\centering \includegraphics[scale=.85] {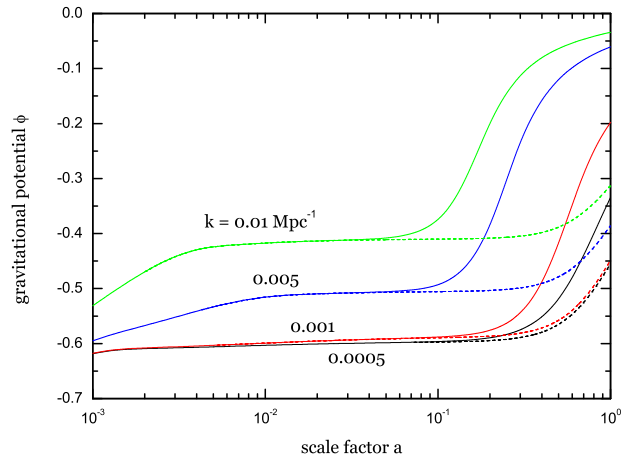}
\caption{(Color Online) The evolution of the gravitational potential $%
\protect\phi $ in the bulk viscosity model (solid curve) as compared to the
results for the standard $\Lambda $CDM paradigm (dashed curves). The results
for 4 different length scales ($k=0.0005$, $0.001,0.005$ and $0.01$ $Mpc^{-1}
$) are shown. Clearly $\protect\phi $ decays much faster in the bulk
viscosity model. }
\label{fig:Figure4}
\end{figure}

\begin{figure}[tbp]
\centering \includegraphics[scale=.85] {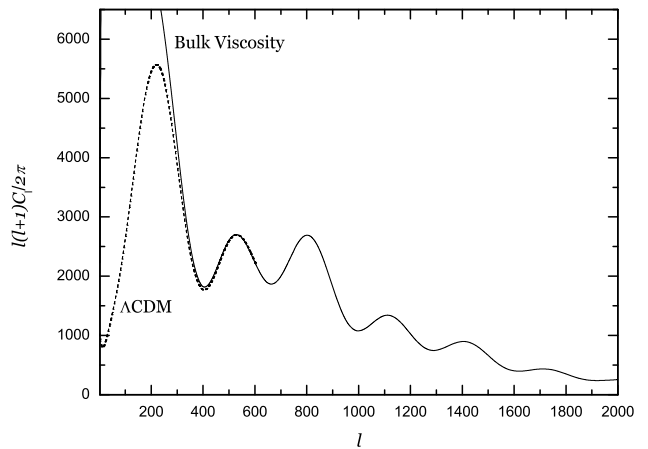}
\caption{The CMB temperature fluctuation spectrum of the bulk viscosity
model (the solid curve) as compared to the predictions of the $\Lambda $CDM
paradigm (the dashed curve). The model parameters are chosen to be $m=-0.4$
and $\protect\beta =0.236$, and all other parameters are the same in the two
models.}
\label{fig:Figure5}
\end{figure}

The  analysis above is generalizable to an arbitrary choice of the viscosity
function $\eta (\rho _{\mathrm{D}})$. To do this, we just need to define
\begin{eqnarray}
\tilde{m}(a) &\equiv&
\frac{\rho_{\mathrm{D}}}{\eta(\rho_{\mathrm{D}})}
\frac{\partial\eta(\rho_{\mathrm{D}})}{\partial\rho_{\mathrm{D}}}
\end{eqnarray}
and replace the constants $m$ in Eqs.~(\ref{eq:C3}, \ref{eq:C4}) with $%
\tilde{m}(a)$. Then, Eq.~(\ref{eq:a_evolve_clxc}) describes the evolution of
the dark matter density contrast in the general case of $\eta (\rho _{%
\mathrm{D}})$.

There are a couple of points to be noted about Eq.~(\ref{eq:a_evolve_clxc}).
First, it is clear that $m=0$ does not guarantee that $C_{2}$ and $C_{4}$
are zero, and the EoS of the viscous dark matter $w$ is important. Since $%
w\neq 0,$ if the viscous dark matter is responsible for accelerating the
expansion of the universe, our conclusion that the evolution of dark matter
density contrast will be sensitively scale-dependent and deviate
significantly from $\Lambda $CDM will hold in general. Note that this is
different from the analysis of Ref.~\cite{Colistete2007}, which considers
the special case of $m=0$ and concludes that the matter power spectrum in
this model behaves well. Second, Eq.~(\ref{eq:a_evolve_clxc}) is
qualitatively different from its counterparts in other models which unify
dark matter and dark energy, such as the (generalized) Chaplygin gas model.
In the latter we have generally $C_{2}=0,$ and so the equation describes an
under-damped oscillator rather than an over-damped one. Consequently, the
dark matter density contrast oscillates with (rapidly) decaying amplitude
(or blows up if $C_{4}$ is negative), in contrast to the monotonic decay in
our model here. Tracing the derivation of Eq.~(\ref{eq:a_evolve_clxc}), it
is easy to find that the $C_{2}$ term comes from the term with $\mathcal{Z}$
in Eq.~(\ref{eq:a_clxcp}), which is created by the fact that $p_{\mathrm{D}%
}\propto \nabla ^{a}u_{a}$. We also note that Eqs.~(\ref{eq:C1}, \ref{eq:C2}%
) are independent of $m$ and hence independent of the functional form of $%
\eta (\rho _{\mathrm{D}})$. As long as $C_{4}>0$, we will obtain a
qualitative picture similar to the one given in Fig.~\ref{fig:Figure3}.

Although the very different evolution of $\Delta _{\mathrm{D}}$ from that in
$\Lambda $CDM is not directly reflected in the observed galaxy power
spectrum, it will modify the gravitational potential and subsequently affect
observables such as the CMB, weak lensing and CMB-galaxy cross-correlation.
As the viscous dark matter contributes the majority of the total energy in
the universe, a decay in its density contrast as in Fig.~\ref{fig:Figure3}
will also drive the gravitational potential $\phi $ to decay significantly.
This point is verified in Fig.~\ref{fig:Figure4}, which clearly shows that
the decay of $\phi $ is much faster than that in the $\Lambda $CDM model.

The fast decay of $\phi $ will enhance the integrated Sachs-Wolfe (ISW)
effect, which then contributes a source term $\propto \int^{\tau _{0}}\phi
^{\prime }j_{\ell }[k(\tau _{0}-\tau )]d\tau $ to the CMB fluctuations,
where $j_{\ell }(k\tau )$ is the spherical Bessel function and $\phi
^{\prime }$ the (conformal) time derivative of $\phi $. This corresponds to
more power in the CMB spectrum on large scales \cite{Comment1}, as displayed
in Fig.~\ref{fig:Figure5}. Note that the positions of the acoustic peaks for
the two models are the same in Fig.~\ref{fig:Figure5}, because their
background evolutions are almost indistinguishable. We can see that
comparison with CMB data could provide strong restrictions on the present
model as well. Again, because the damping in $\Delta _{\mathrm{D}}$ is a
general feature in UDM models based on bulk viscosity, so are the
enhancements of the ISW effect and the low-$\ell $ CMB power. In principle,
the weak-lensing convergence power spectrum, which reflects the (projected)
potential distribution along the line of sight, will also be modified (as
compared to the $\Lambda $CDM result) significantly by the rapid decay of
the potential. This is not considered here because the CMB spectrum itself
already places stringent constraint on the model.

\section{Discussion and Conclusions}

\label{sect:conclusion}

To summarize, in this work we have investigated the background cosmological
evolution and large-scale structure formation in the bulk viscous models
designed as alternatives to dark energy. A bulk viscosity generates an
effective pressure $p=-3\eta (\rho )H$ \ which, when the function $\eta
(\rho )$ is appropriately chosen, could drive the acceleration of the cosmic
expansion. Our numerical calculation focuses on a particular choice $\eta
(\rho )=\alpha \rho ^{m}$ with $\alpha ,m$ being constants. Such a choice
has been considered before in the literature \cite{Barrow1988}, in the
contexts of both inflationary and late-time accelerating universes. It is
interesting to note that with suitable values of $m$, the viscous matter
behaves as cold dark matter at early times and as a cosmological constant in
the future; thus this model naturally unifies dark matter and dark energy,
at least at the background level.

We have used the measurements of supernovae luminosity distance
and CMB shift parameter to constrain the model parameters and
found that for the
best-fitting parameters, $m=-0.4$ and $\beta =0.236$ ($\beta \propto \alpha $%
), the background evolution of the universe is almost the same as that
predicted by the $\Lambda $CDM model. From this viewpoint it seems that the
model is a feasible alternative to an explicit cosmological constant.

However, when it comes to the linear perturbations, the viscous dark matter
starts to behave very differently from cold dark matter. The pressure of the
viscous dark matter tends to resist the growth of the density contrast, and
when it becomes significant (at late times) this effect can rapidly damp out
the density perturbations, particularly on small scales. Even though this
smoothing of the viscous dark matter density perturbation cannot be seen
directly, it could reduce the growth rate of density perturbations in the
luminous matter (which can be measured by the galaxy power spectrum) and
drive the fast decay of the gravitational potential fluctuations, which
subsequently modifies the large-scale CMB spectrum, weak lensing and
CMB-galaxy cross-correlations.

Note that the model we consider has no explicit $\Lambda $CDM limit, \emph{%
i.e.}, one cannot adjust the model parameters $\alpha $ and $m$ to make the
model reduce to $\Lambda $CDM exactly, unless a cosmological constant is
added and the limit $\alpha \rightarrow 0$ is taken. The latter case, in
which we have both an explicit cosmological constant and viscous matter, is
not particularly appealing because it will introduce more complexity without
solving any of the problems of $\Lambda $CDM. Rather, we are interested in
whether the viscous dark matter alone could replace $\Lambda $CDM
completely. This means that the EoS parameter $w$ will necessarily be of
order $-1$ at late times. According to our analysis in \S ~\ref%
{sect:PerturbedEvolution} [cf.~Eqs.~(\ref{eq:a_evolve_clxc}-\ref{eq:C4})],
it is the value of $w$ that determines the evolution of $\Delta _{\mathrm{D}}
$, and so we expect that all the qualitative pictures given in \S~\ref%
{sect:PerturbedEvolution} will remain in place in any attempts to replace
dark matter and dark energy with bulk viscous matter alone. Viscous matter
is therefore not a successful contender for the dark sector.

The bulk viscosity model is another candidate to explain the dark energy
without introducing dynamical degrees of freedom. In order to explain the
accelerating cosmic expansion, one needs a negative (effective) pressure. If
no new dynamical degree of freedom is introduced, the negative pressure
should be either a constant, or a function of $\rho _{\mathrm{DE}}$ (the
energy density of the newly added matter), $\rho _{m}$ (the energy density
of the existing matter) or $H$, or combinations of these variables, which
are all the possible variables in the background cosmology \cite%
{Comment2} (see \cite{cuscuton, cuscutonb} for a counterexample
however, where a new degree of freedom is added but is made
non-dynamical). Hence, we have the following conclusions for each
of the allowed prescriptions for a form of newly added matter
\cite{Nojiri2005a, Nojiri2005b, Nojiri2005c}:

\begin{enumerate}
\item The case $p=\mathrm{const.}$ is simply a cosmological constant.

\item An example of the case $p=p(\rho _{\mathrm{DE}})$ is the (generalized)
Chaplygin gas model \cite{Kamenshchik2001, Bento2002}. The large-scale
structure formation here has been considered in \cite{Sandvik2004, Bean2003,
Carturan2003}. In this class of models we have $\mathcal{X}^{p}\propto
\mathcal{X}$ in general, and this leads to a term proportional to $%
k^{2}\Delta $ in the evolution equation for the density contrast $\Delta $
of the \emph{newly added matter}, and this term dominates on small scales ($%
k\gg \mathcal{H}$). Depending on the sign of this new term, $\Delta $ will
either blow up or oscillate and decay rapidly on small scales.

Note that this class of models are frequently considered as either
UDM models or dark energy models. In the former case, as shown in
\cite{Sandvik2004}, the constraints on the models are particularly
stringent (note however that, as pointed out in \cite{Avelino2004,
Beca2007}, in these models the nonlinear corrections can be
important, which will make the simple linear treatments inaccurate
or even incorrect, and will potentially significantly modify the
background evolution as well, depending on the parameters used).
In the latter case, density perturbations of CDM and baryons may
not be affected significantly, much like the fact that in the
above viscous model the baryons are not greatly affected. However,
there generally will be other distinct new features of the model
\cite{Bean2003}.

\item Examples for the case $p=p(\rho_{m})$ include the Cardassian model
\cite{Gondolo2003}, the Palatini $f(R)$ gravity \cite{Vollick2003} and the $%
\omega =-3/2$ Brans-Dicke theory with a potential, the linear
perturbations of which have been considered in \cite{Koivisto2005,
Koivisto2006, Li2006, Li2007, Li2007b}. In these models, by a
similar argument as in the above, there will be a term
proportional to $k^{2}\Delta_{m}$ in the evolution equation for
the \emph{existing matter} density perturbation $\Delta_{m}$. As a
result, $\Delta_{m}$ will experience blowing-up or rapidly
decaying oscillation on small scales at late times. Note, however,
that in this class of models the averaging over the microscopic
structure of the (existing) matter distribution may be a serious
issue, rendering the appropriately averaged cosmological behaviour
very different from naive predictions \cite{Li2008a, Li2008b} (in
a way similar to \cite{Beca2007} though with some differences,
namely in \cite{Li2008a, Li2008b} the averaging is over
microscopic scales while in \cite{Beca2007} it is over
astronomical scales). This is because $p$ for the newly added
matter depends algebraically on the density of normal matter
particles, and could be very different inside and outside the
distribution of the latter. As the normal matter particles only
occupy a tiny portion of the total volume of the universe, after
averaging, $p$ should be dominated by its value outside the normal
matter particles (\emph{i.e.}, in the vacuum), which is likely to
be a constant.

\item The case $p=p(H)$ includes bulk viscosity model with $\eta (\rho )=%
\mathrm{const.}$ considered in \cite{Colistete2007, Avelino2008a,
Avelino2008b, Avelino2008c}, which is a special case for the general model
we considered in this work, with $p=p(\rho _{\mathrm{DE}},H)$. Here, the
dependence of $p$ on $H$ results in the evolution equation of $\Delta $ for
the \emph{newly added} matter acquiring both a term proportional to $%
k^{2}\Delta $ and one proportional to $k^{2}\Delta ^{\prime }$.
Consequently, on small scales at late times, $\Delta $ will decay
rapidly without oscillation like an over-damped oscillator. We
have shown that the UDM model based on this is generally unable to
pass several cosmological tests. Note that as in the case of
$p=p(\rho_{\mathrm{DE}})$, the averaging issue again needs to be
taken into account if more precise predictions are to be obtained
\cite{Avelino2004, Beca2007}, though its full effect and
significance need careful nonlinear studies such as N-body
simulations.
\end{enumerate}

\ \ \ \ \ \ In all such cases (except $p=\mathrm{const.}$), we have seen
that the evolution of $\Delta $ (for either the existing matter or the newly
added matter) becomes very irregular. If the matter with irregular
perturbation evolution makes a significant contribution to the total energy
of the universe, then this will lead to large modifications to the $\Lambda $%
CDM predictions for the matter power spectrum, CMB, weak lensing
and similar tests. This indicates that a viable dynamical UDM
model is likely to involve extra dynamical degrees of freedom in
contrast to that provided by a (generalized) Chaplygin gas or bulk
viscosity \cite{Bertacca2007a, Bertacca2007b, Bertacca2008a,
Bertacca2008b}.

\

\begin{acknowledgments}
We thank Niayesh Afshordi for kindly pointing out
Ref.~\cite{cuscuton, cuscutonb} to us. BL acknowledges financial
supports from an Overseas Research Studentship, the Cambridge
Overseas Trust, the DAMTP and Queens' College at Cambridge.
\end{acknowledgments}


\begin{thebibliography}
\bibitem[Comment(2008)]{Comment1} {The other contribution to the enhanced
large-scale CMB power comes from the photon density perturbation, which is
also affected by the evolution of $\Delta_{\mathrm{D}}$.}
\bibitem[Comment(2008)]{Comment2} {If $p=p(\dot{\rho}_{m}, \dot{\rho}_{%
\mathrm{DE}})$, then using the conservation equation we can reexpress $p$ as
$p=p(\rho_{m}, \rho_{\mathrm{DE}}, H)$. Further possibilities $p=p(\dot{H},
\cdots)$ are too complex to be considered here, but by the same logic as
used in deriving Eq.~(\ref{eq:a_evolve_clxc}) we could show that these will
also lead to scale-dependent evolution of density perturbation. For example,
if $p=p(\rho, H, \dot{H})$ then we can find that there are terms
proportional to $k^{2}$ in the coefficients of $\Delta$, $\Delta^{\prime }$
and also $\Delta^{\prime \prime }$.}
\end{thebibliography}
\end{document}